\documentstyle[12pt]{article}

\makeatletter
\@addtoreset{equation}{section}
\makeatother

\def\be{\begin{equation}}       \def\eq{\begin{equation}}
\def\ee{\end{equation}}         \def\eqe{\end{equation}}

\def\bea{\begin{eqnarray}}      \def\eqa{\begin{eqnarray}}
\def\ena{\end{eqnarray}}        \def\eea{\end{eqnarray}}
                                \def\eqae{\end{eqnarray}}

\def\ba{\begin{array}}
\def\ea{\end{array}}
\def\unit{1 \hskip-.3em \raise2pt\hbox{$ \scriptstyle |$ } }
\def\b{\beta}
\def\c{\gamma}
\def\d{\delta}
\def\e{\epsilon}           % Also, \varepsilon
\def\f{\phi}               %      \varphi
\def\vf{\varphi}  \def\tvf{\tilde{\varphi}}
\def\g{\gamma}

\def\k{\kappa}                    % Also, \varkappa (see below)
\def\l{\lambda}
\def\m{\mu}
\def\n{\nu}
  
\def\p{\pi}                % Also, \varpi
  \def\th{\theta}                  %     \vartheta
\def\r{\rho}                                     %     \varrho
\def\s{\sigma}                                   %     \varsigma

\def\x{\xi}

\def\D{\Delta}
\def\F{\Phi}
\def\G{\Gamma}

\def\L{\Lambda}
\def\O{\Omega}

\def\ca{{\cal A}}

\def\cf{{\cal F}}

\def\co{{\cal O}}

\def\car{{\cal R}}

\def\half{{1 \over 2}}

\def\bop#1{\setbox0=\hbox{$#1M$}\mkern1.5mu
        \vbox{\hrule height0pt depth.04\ht0
        \hbox{\vrule width.04\ht0 height.9\ht0 \kern.9\ht0
        \vrule width.04\ht0}\hrule height.04\ht0}\mkern1.5mu}
\def\Box{{\mathpalette\bop{}}}                        % box
\def\pa{\partial}                              % curly d
\def\de{\nabla}                                       % del
\def\>{\rangle} %right angle

\def\<{\langle} %left angle
\def\Dsl{D \hskip-.6em \raise1pt\hbox{$ / $ } }

\def\leftrightarrowfill{$\mathsurround=0pt \mathord\leftarrow \mkern-6mu
       \cleaders\hbox{$\mkern-2mu \mathord- \mkern-2mu$}\hfill
       \mkern-6mu \mathord\rightarrow$}
\def\dvec#1{\vbox{\ialign{##\crcr
       \leftrightarrowfill\crcr\noalign{\kern-1pt\nointerlineskip}
       $\hfil\displaystyle{#1}\hfil$\crcr}}}          % <--> accent
\def\hook#1{{\vrule height#1pt width0.4pt depth0pt}}
\def\leftrighthookfill#1{$\mathsurround=0pt \mathord\hook#1
       \hrulefill\mathord\hook#1$}
\def\underhook#1{\vtop{\ialign{##\crcr                 % |_| under
       $\hfil\displaystyle{#1}\hfil$\crcr
       \noalign{\kern-1pt\nointerlineskip\vskip2pt}
       \leftrighthookfill5\crcr}}}
\def\smallunderhook#1{\vtop{\ialign{##\crcr      % " for su'scripts
       $\hfil\scriptstyle{#1}\hfil$\crcr
       \noalign{\kern-1pt\nointerlineskip\vskip2pt}
       \leftrighthookfill3\crcr}}}
\def\to{\rightarrow}

\def\pa{\partial}

\newcommand{\tnnn}[1]{\mbox{\tiny #1}}
\def\GN{G_{\mbox{\tnnn N}}}
\def\de{\mbox{d}}

\def\nonu{\nonumber \\{}}
\def\half{{1 \over 2}}
\def\Tr{{\rm Tr}\, }

\begin{document}
%\null
%\vskip-60pt

\begin{flushright}
PUTP-2047\\
{\tt hep-th/0209067}
\end{flushright}
\vskip0.3truecm

\begin{center}
\vskip 1truecm
\begin{center}
{\Large\bf
%\titleline
Lecture Notes on \\
Holographic Renormalization}
\end{center}
\vskip 1truecm
%\vfill
{\large\bf
Kostas Skenderis\footnote{{\tt
kostas@feynman.princeton.edu}}} \\
\vskip 0.5truemm
{\it Physics Department, Princeton University \\
Princeton, NJ 08544, USA}
\end{center}
\vskip .5truecm
\begin{abstract}
\vskip .2truecm
We review the formalism of holographic renormalization.
We start by discussing mathematical results on asymptotically 
anti-de Sitter spacetimes. We then outline the general method of holographic 
renormalization. The method is illustrated by working 
all details in a simple example: a massive scalar field on
anti-de Sitter spacetime. The discussion includes 
the derivation of the on-shell renormalized action, of 
holographic Ward identities, anomalies and RG equations, 
and the computation of renormalized one-, two- and four-point functions.
We then discuss 
the application of the method to holographic RG flows.
We also show that the results of the 
near-boundary analysis of asymptotically AdS spacetimes
can be analytically continued to apply to asymptotically 
de Sitter spacetimes. In particular, it is shown that 
the Brown-York stress energy tensor of de Sitter spacetime
is equal, up to a dimension dependent sign, to the Brown-York 
stress energy tensor of an associated AdS spacetime.

\end{abstract}

\vfill
\pagebreak
\setcounter{page}{1}

\tableofcontents
\addtocontents{toc}{\protect\setcounter{tocdepth}{2}}

\newpage

\section{Introduction}

The AdS/CFT correspondence 
\cite{Maldacena:1997re,Gubser:1998bc,Witten:1998qj,Aharony:1999ti,D'Hoker:2002aw} 
offers the best understood example 
of a gravity/gauge theory duality. According to this duality 
string theory on an asymptotically anti-de Sitter spacetime (AAdS) times
a compact manifold $M$ is exactly equivalent to the 
quantum field theory (QFT) ``living'' on the boundary of AAdS. 
This is a strong/weak coupling duality:
the strong coupling regime of the quantum field theory 
corresponds to the weak coupling regime of the string theory 
and vice versa. The exact equivalence between the two formulations
means that, at least in principle, one can obtain complete 
information on one side of the duality by performing 
computations on the other side. In these lecture notes
we discuss how to obtain renormalized QFT correlation functions by 
performing computations on the gravity side of the correspondence.

Despite much effort, string theory on AdS spacetimes
is still poorly understood. At low energies, however,
the theory is well approximated by supergravity. 
The relevant description is in terms of a $d$+1
dimensional supergravity theory with an AAdS ground state 
coupled to the Kaluza-Klein (KK) modes that result
from the reduction of the 10$d$ (or 11$d$) supergravity 
on the compact manifold $M$. On the gauge theory side
the low energy limit corresponds to considering the 
large 't Hooft limit, $\l=g_{YM}^2 N >> 1$. 
Suppressing loop contributions on the gravitational side
corresponds to considering the $QFT$ in the large $N$ 
limit. In this limit the gravitational computations
involve classical solutions of the supergravity 
theory coupled to the infinite set of the KK modes.
In some cases one can further simplify things by 
consistently truncate the KK modes.
In these cases it is sufficient to work with 
a lower dimensional gauged supergravity.
The existence of a 
consistent truncation implies that there is a subset of 
gauge invariant operators on the gauge theory side
which are closed under OPE's. To compute correlation
functions of these operators it is sufficient to 
study classical solutions of the $d+1$ dimensional 
gauged supergravity.

In quantum field theory, correlation functions suffer from UV divergences,
and one needs to renormalize the theory to make sense of them.
A general phenomenon of the gravity/gauge theory correspondence
is the so-called UV/IR connection \cite{Susskind:1998dq}, 
i.e. UV divergences in 
the field theory are related to IR divergences on the 
gravitational side, and vice versa. On the gravitational
side, long distance (IR) is the same as near the boundary.
The purpose of these lecture notes is explain how to 
deal with these IR divergences, i.e. how to 
``holographically renormalize''.

In the QFT the cancellation of the UV divergences does not 
depend on the IR physics. This implies that the 
holographic renormalization should only depend on 
the near-boundary analysis. Furthermore, in QFT 
a major role in the renormalization program is 
played by symmetries and the corresponding Ward 
identities. If the UV subtractions respect some 
symmetry then the corresponding Ward identity holds, otherwise
it is anomalous. The corresponding statement 
on the gravitational side is that the near-boundary 
analysis that determines the IR divergences should be sufficient to establish 
the holographic Ward identities and anomalies.
On the other hand, correlation functions capture 
the dynamics of the theory so the near-boundary 
analysis should not be sufficient to determine them.
To determine correlation functions one 
needs exact solutions of the bulk field equations.
The subtractions necessary to render the correlations
functions finite should be consistent with each 
other. This is guaranteed if they are made by 
means of counterterms.  In early studies of 
correlation functions in AdS/CFT the divergences
were simply dropped from correlators. In the
holographic renormalization program
we shall implement all subtractions by means of 
covariant counterterms. 

The regularization and renormalization method discussed
in these notes was first introduced in \cite{HS}
and it was promoted to a systematic method 
in \cite{dHSS}. It was applied to RG flows 
in \cite{howtogo,holren}. Counterterms for AdS 
gravity were also introduced in \cite{Balasubramanian:1999re},
see also \cite{Myers,EJM,Mann:1999bt,KLS}.
The holographic renormalization method is described in 
detail in \cite{holren}. The discussion in these
notes should be viewed as complementary to the discussion 
there. In general we refrain from discussing in detail
material that are sufficiently discussed elsewhere. 
The emphasis here is in the general features of the 
method illustrated by the simplest possible example.
 
These lecture notes are organized as follows.
In the next section we state the main result,
namely the reformulation of the computation of 
correlation function in terms of renormalized
1-point functions in the presence of sources.
In section \ref{aads} we introduce
asymptotically AdS spacetimes and we discuss 
in detail the results of Fefferman and Graham
\cite{FeffermanGraham} on hyperbolic manifolds.
In section \ref{HRM} we present the method 
of holographic renormalization. In particular, 
we discuss how to obtain asymptotic solutions
of the field equations, 
regularize and renormalize the on-shell action,
compute exact 1-point functions, derive 
Ward identities, compute correlation functions 
and derive RG equations. All of these
are illustrated by a simple example, a massive
scalar in AdS, in section \ref{msc}. In section \ref{rgsec}
we discuss the application of the method to holographic
RG flows. We conclude by discussing open problems and future
directions. 

In the appendix we show that all {\it local} results derived
via the near-boundary analysis can be straightforwardly 
``analytically continued'' to asymptotically dS spacetimes.
In particular, the Brown-York stress energy tensor \cite{BrownYork}
associated with an asymptotically dS spacetime \cite{BBM}
is always equal, up to a sign, to the stress energy 
of the AdS space related to the dS space 
by a specific analytic continuation. The sign is plus 
in $D=4 k + 3$ (bulk) dimensions and minus in 
$D=4 k + 1$ dimensions. In particular, in three dimensions
the two stress energy tensors are the same and in 
five dimensions they differ by a sign. 

\section{Statement of results}

According to the gravity/gauge theory correspondence, for every 
bulk field $\Phi$ there is a corresponding gauge invariant 
boundary operator $O_\F$. In particular, bulk gauge 
fields correspond to boundary symmetry currents. 
As we will review later, AAdS spaces have a boundary 
at spatial infinity, and one needs to impose 
appropriate boundary conditions there. The partition
function of the bulk theory is then a functional
of the fields parametrizing the boundary values 
of the bulk fields. According to the prescription
proposed in \cite{Gubser:1998bc,Witten:1998qj},
the boundary values of the fields are identified 
with sources that couple to the dual operator,
and the on-shell bulk partition function with the 
generating functional of QFT correlation functions,
\be
Z_{{\rm SUGRA}}[\phi_{(0)}] = 
\int_{\Phi \sim \f_{(0)}} D \Phi  \exp(-S[\Phi]) 
= \< \exp \left( -\int_{\pa AAdS} \phi_{(0)} O \right) \>_{QFT}.
\ee
where the expectation value on the right hand side is 
over the QFT path integral, and $\pa AAdS$ denotes the 
boundary of the asymptotically AdS space.
We will work exclusively in the leading 
saddle point approximation where this relation 
becomes,
\be \label{GKPW}
S_{{\rm onshell}}[\phi_{(0)}] = - W_{QFT}[\phi_{(0)}]
\ee
where $S_{{\rm onshell}}[\phi_{(0)}]$ is the on-shell
supergravity action and 
$W_{QFT}[\phi_{(0)}]$ is the generating
function of QFT connected graphs.
 Correlation functions of the operator $O$ are now computed
by functional differentiation with respect to the source,
\bea \label{npt}
\<O(x)\> &=& 
{\delta S_{{\rm onshell}} \over \delta \phi_{(0)}(x)} \Big|_{\f_{(0)}=0} \nonu
\<O(x) O(x_2) \> &=& -
{\delta^2 S_{{\rm onshell}} \over 
\delta \f_{(0)} (x) \d \phi_{(0)}(x_2)} \Big|_{\f_{(0)}=0} 
\nonu
\<O(x_1) \cdots O(x_n) \> &=& (-1)^{n+1} {\delta^3 S_{{\rm onshell}} \over 
\delta \f_{(0)} (x_1) \cdots \d \phi_{(0)}(x_n)} \Big|_{\f_{(0)}=0} 
\eea
etc.  

As we mentioned, QFT correlation functions diverge and the 
right hand side of (\ref{GKPW}) is not well-defined 
without renormalization. Similarly, the left hand side
is also divergent due to the infinite volume
of the spacetime and one needs to appropriately renormalize.
In section \ref{HRM} we will lay out the renormalization method
but we start by stating the final result.

Given a classical  action $S[\Phi, A_\mu, G_{\m \n}, ...]$  
that depends on a number of fields $\Phi, A_\mu, G_{\m \n}$, etc
there exist {\it exact} renormalized 1-point functions, one
for each bulk field,
\bea \label{1pt}
\Phi \to \< O(x) \>_s  = {1 \over \sqrt{g_{(0)}(x)}} 
{\delta S_{{\rm ren}} \over \d \f_{(0)}(x)} 
\sim \f_{(2 \D - d)}(x) \nonu
A_\mu \to \< J_i(x) \>_s ={1 \over \sqrt{g_{(0)(x)}}}  
{\delta S_{{\rm ren}} \over \d A_{i (0)}(x)} 
\sim A_{{m}i}(x) \nonu 
G_{\m \n} \to \< T_{ij}(x) \>_s = {2 \over \sqrt{g_{(0)(x)}}}  
{\delta S_{{\rm ren}} \over \d A_{i (0)}(x)} \sim g_{(d) ij}(x) 
\eea
where $S_{{\rm ren}}$ is the renormalized on-shell action
(to be discussed shortly),
$J_i$ is the boundary symmetry current that couples
to the bulk gauge field $A_\mu$, $T_{ij}$ is the boundary 
stress energy tensor that couples to the boundary metric
$g_{(0)ij}$ etc. The fields on the right hand side,
$\f_{(2\D-d)}, A_{{m}i}, g_{(d) ij}$ appear in the asymptotic 
expansion of the solutions of the bulk field equation.
The exact definition is given in the section \ref{HRM}. 
The asymptotic analysis of the field equations does 
not determine these coefficient but given
an exact solution of the field equations 
it is straightforward to extract $\f_{(2\D-d)}, A_{{m}i}, g_{(d) ij}$.  
These functions are in general non-local functions
of the sources $\f_{(0)},  A_{{0}i}, g_{(0) ij}$.
Notice that the relations (\ref{1pt}) hold for {\it any} solution of the 
bulk field equations,
and as such it is a property of the theory rather than of 
each specific solution.
 
The subscript $s$ in (\ref{1pt}) denotes that the expectation
values are in the presence of sources. This means that if we 
want to compute higher point functions we only need to 
differentiate (\ref{1pt}) and then set the sources to zero,
\be \label{rnpt}
\< O(x_1) \cdots O(x_n) \> 
\sim {\d \f_{(2 \D -d)}(x_1) \over \d \f_{(0)}(x_2) ... 
\d \f_{(0)}(x_n)} \Big|_{\f_{(0)}=0}
\ee
This relation replaces the relations (\ref{npt}). 
If one knew $\f_{(2 \D -d)}$ exactly as a function of sources, 
one would have solved the theory since using (\ref{rnpt})
one can determine all $n$-point functions. We will see
later how to determine $\f_{(2 \D -d)}$ using (bulk) perturbation
theory.

The exact one-point functions allow one to establish 
the holographic Ward identities in full generality.
For example, in the case where the only source turned on 
is the boundary metric, i.e. we only consider the 
Einstein-Hilbert term in the bulk action, one can show that 
the bulk field equations imply the correct
diffeomorphism and conformal Ward identities \cite{dHSS},
\bea
\nabla^i \<T_{ij} \>_s &=& 0 \nonu
\<T^i_i \>_s &=& \ca \nonu
\eea
where $\ca$ is the holographic Weyl anomaly \cite{HS}.

\section{Asymptotically anti-de Sitter spacetimes} \label{aads}

In this section we discuss in detail asymptotically AdS spacetimes.
Recall that the AdS spacetime is a maximally symmetric solution of Einstein's
equations with negative cosmological constant,
\be \label{einstein}
R_{\m \n} - \half R G_{\mu \nu} = \L G_{\mu \nu}
\ee
The AdS space is conformally flat. This implies that the Weyl tensor
vanishes, 
\be \label{weyl}
W_{\mu \nu \k \l}=0.
\ee
This equation combined with Einstein equations 
implies that the curvature tensor of the $AdS_{d+1}$ spacetime 
is given by\footnote{Our curvature convention is \label{conventions}
$R_{\m \n \k}{}^\l = \pa_\m \G_{\n \k}^\l + \G_{\m \r}^\l \G_{\n \k}^\r
- \m \leftrightarrow \n, R_{\m \n}= R_{\m \l \n}{}^\l, R=G^{\m \n} R_{\m \n}$.}
\be \label{curAdS}
R_{\m \n \k \l} = {1 \over l^2} 
(G_{\k \m} G_{\n \l} - G_{\m \l} G_{\n \k})
\ee
where $l^2$ is the AdS radius ($\L=-d(d-1)/2 l^2$).

The metric (in convenient coordinates) is given by 
\be
ds^2 = {l^2 \over \cos^2 \th} (-dt^2 + d \th^2 + \sin^2 \th d \O_{d-1}^2)
\ee
where $0 \leq \th < \p/2$.
Notice that the metric has a second order pole at $\theta=\pi/2$.
This is where the boundary of AdS is located. Because of the 
second order pole, the bulk metric does not yield a metric at the 
boundary. It yields a conformal structure instead.
Let us consider a function $r(x)$ that is positive in the 
interior of AdS, but has a first order pole at the boundary.
Such a function is called a ``defining function''.
We now multiply the AdS metric by $r^2$ and evaluate it at the 
boundary,
\be
g_{(0)} = r^2 G|_{\pi/2}
\ee
For instance, one could choose $r=\cos \th$.
This metric is finite but it is only defined up to conformal transformations.
Indeed, if $r$ is a good defining function, then so is $re^w$,
where $w$ is a function with no zeros or poles at the boundary. So 
the AdS metric yields a conformal structure at the boundary,
i.e. a metric up to conformal transformations.

We will now define asymptotically AdS spaces by generalizing 
the above considerations. First, let us define {\it conformally
compact manifolds} following \cite{penrose}. Let $X$ be the interior of 
a manifold-with-boundary $\bar{X}$, and let $M=\pa X$ be its 
boundary. We will call a metric $G$ conformally compact
if it has a second order pole at $M$ but there exists a 
defining function (i.e. $r(M)=0$, $dr(M) \neq 0$ and $r(X)>0$)
such that 
\be \label{gdef}
g = r^2 G
\ee
smoothly extends to $\bar{X}$, $g|_M = g_{(0)}$, and is 
non-degenerate.
As in the case of AdS space, this procedure defines a
conformal structure on $M$. There is another quantity 
that smoothly extends to $\bar{X}$ and its restriction 
to $M$ is independent of choices (i.e. it depends only 
on the conformal structure),
\be
|dr|_g^2 = g^{\m \n} \pa_\m r \pa_\n r
\ee
This can be shown by using the definition (\ref{gdef}).

One can calculate the curvature of the bulk metric, $G$,
\be
R_{\k \l \m \n}[G] =  |dr|_g^2 (G_{\k \m} G_{\n \l} - G_{\m \l} G_{\n \k})
 + \co(r^{-3})
\ee
Notice that the leading term is of order $r^{-4}$.
So conformally compact manifolds have a curvature tensor
that near to the boundary (i.e. $r=0$) looks like the curvature 
tensor of AdS space (\ref{curAdS}). Notice that up to this point we did not 
impose that the metric $G$ is Einstein, i.e. satisfies (\ref{einstein}).
A short computation shows that Einstein's equations imply,
\be
|dr|_g^2|_M = {1 \over l^2}
\ee
In this case the Riemann tensor near the boundary is exactly the 
same as that of AdS space. We are thus lead to the 
following definition \newline
{\bf Definition:} {\it An Asymptotically AdS metric is a conformally 
compact Einstein metric.} \newline
We set $l=1$ from now on.
Notice that this definition does not impose any condition 
on the topology of the boundary.

A question of interest for us is whether given a conformal structure 
at infinity one can determine an Einstein bulk metric with the prescribed
boundary conditions. This has been answered in 
\cite{FeffermanGraham} (see \cite{Graham} for a review).
The first step is to prove the following theorem \newline 
{\it Theorem:} There is always a preferred defining function
such that 
\be \label{dr}
|d r|_g^2 = 1
\ee
in a neighborhood of the boundary $M$. \newline
The idea of the proof is as follow.
Let $r_0$ be a defining function such that (\ref{dr})
does not hold, and consider another defining function 
$e^w r_0$. Then equation (\ref{dr}) becomes a 
differential equation for $w$ that always has a solution 
in the neighborhood of $M$ (recall that $|dr|_g^2|_M = 1$).

We now consider Gaussian coordinates emanating from the 
boundary. We take the inward (radial) coordinate to be 
the affine parameter of the geodesics with tangent
$\nabla r$. Clearly these are good coordinates as long 
as we do not meet any caustics. In particular, we can 
take the defining function as the radial coordinate.
Then the bulk metric in the neighborhood of the boundary
takes the form,
\be
ds^2 = {1 \over r^2} (dr^2 + g_{ij}(x,r) dx^i dx^j)
\ee
By construction, $g_{ij}(x,r)$ has a smooth limit 
as $r \to 0$, so it can be written as
\be
g_{ij}(x,r) = g_{(0)ij} + r g_{(1) ij} + r^2 g_{(2)ij} + ...
\ee
One may now determine the coefficients $g_{(k) ij}, k>0$
from Einstein's equations. Explicit computation shows that 
in pure gravity  all coefficients multiplying odd powers of 
$r$ vanish up  to the order $r^d$. To simplify the computation
of the even coefficients we introduce the new coordinate
$\r=r^2$ \cite{HS}. This is the coordinate used throughout this
paper. In these coordinates the metric is given by
\bea \label{coord}
&&ds^2=G_{\m \n} dx^\m dx^\n = {d\r^2 \over 4 \r^2} + 
{1 \over \r} g_{ij}(x,\r) dx^i dx^j, \nonu
&&g(x,\r)=g_{(0)} + \cdots + \r^{d/2} g_{(d)} + h_{(d)} \r^{d/2} \log \r + ... 
\eea
The analysis of the Einstein equations is exactly analogous to the 
analysis of the scalar field equation that we will 
discuss in detail in section \ref{msc}. Details can be 
found in \cite{dHSS}. Here we briefly summarize the main points.
Einstein's equations can be solved order by order in the 
$\r$ variable. The resulting equations are {\it algebraic}
so the solution is insensitive to the sign of the 
cosmological constant and the signature of spacetime.
In appendix A we use this fact to derive the corresponding
asymptotic expansion for asymptotically de Sitter spacetimes.
However, when the cosmological constant is equal to 
zero the corresponding equations are differential \cite{dHSS2}
and impose restrictions on $g_{(0)}$ as well.
This means that, in general, the various coefficients 
in the asymptotic expansion of the metric that contribute
to divergences in the on-shell actions are non-local 
with respect to each other. This implies that in the 
case of asymptotically flat spacetimes there is
no universal set of local counterterms  that can remove the 
divergences from the on-shell action for any solution.
This is one of the main reasons the program of holographic
renormalization does not extend in any straightforward way to the 
case of asymptotically flat spacetimes.

In the case at hand, the equations uniquely determine
the coefficients $g_{(2)},...,g_{(d-2)}$, $h_{(d)}$ 
and the trace and covariant divergence of $g_{(d)}$. 
The coefficient $h_{(d)}$ is present only when $d$ is even,
and it is equal to the metric variation of the holographic
conformal anomaly. The explicit expressions for
$g_{(2)},...,g_{(d-2)},h_{(d)}$ and the trace and covariant 
divergence of $g_{(d)}$ can be found in appendix A of \cite{dHSS}.
$g_{(d)}$ is directly related to the 1-point function of the 
dual stress energy tensor. In general, the solution obtained
by this procedure is only valid near the boundary. 
More powerful techniques are needed in order to obtain solutions
that extend to the deep interior. The three dimensional 
case is special in that one can exactly solve the equations 
to all orders \cite{Skenderis:1999nb}. Even in this case,
however, the coordinate patch (\ref{coord}) does not 
in general cover the entire spacetime, see 
\cite{Rooman:2000ei,Krasnov:2001cu} for related work. 
A review of the purely gravitational case can be found 
in \cite{Skenderis:2000in}.

These results were extended in the case of matter coupled to gravity 
in \cite{dHSS}. In this case the bulk equation reads
\be \label{matterE}
R_{\m \n} - \half R G_{\mu \nu} = T_{\mu \nu}
\ee
where $T_{\mu \nu} = \L G_{\mu \nu} +$ matter contribution.
The equations in this case have a near-boundary solution
provided the matter contribution to $T_{\mu \nu}$ is softer than 
the cosmological constant contribution.
In these cases the matter fields are dual to marginal 
or relevant operators. If the matter stress energy tensor
diverges faster than the cosmological constant term,
the matter fields correspond to irrelevant operators.
In this case in order to obtain a near-boundary solution 
the sources should be considered infinitesimal.

\section{Holographic Renormalization Method} \label{HRM}

In this section we outline all steps involved
in the method of holographic renormalization.

\subsection{Asymptotic solution} 
In the first step we obtain the most general solution 
of the bulk field equations with prescribed, but arbitrary,
Dirichlet boundary condition. Let us suppress all spacetime
and internal 
indices and denote collectively bulk fields by $\cf(x,\r)$.
Near the boundary  each field has an asymptotic expansion of 
the form
\be \label{ansatz}
\cf(x,\r) = \r^m \left(f_{(0)}(x) + f_{(2)}(x) \r + \cdots +
\r^n (f_{(2 n)}(x) + \log \r \tilde{f}_{(2 n)}(x)) + ... \right)
\ee
where $\r$ is the radial coordinate of AdS. 
We use coordinates where the AAdS metric takes the 
asymptotic form,
\bea \label{ads}
ds^2 &=& {d \r^2 \over 4 \r^2} + {1 \over \r} g_{ij}(x,\r)  dx^i d x^j, \nonu
g_{ij}(x,\r) &=& g_{(0)ij}(x) + g_{(2)ij}(x) \r + \cdots
\eea
This is the coordinate system that has been discussed
in section \ref{aads}. The case $g_{(0)ij}=\d_{ij}, g_{(2k)ij}=0, k>0$, 
yields the AdS metric (setting $\r=z^2$ one gets the  
AdS metric in Poincar{\' e} coordinates).\footnote{Throughout these notes 
we work with Euclidean signature. Most of the results, however, are 
independent of the signature of spacetime.}

The field equations are second order differential equations 
in $\r$, so there are two independent solutions. Their 
asymptotic behaviors 
are $\rho^{m}$ and $\rho^{m+n}$, respectively.
In almost all examples discussed in the literature, 
$n$ and $2m$ are non-negative integers and the expansion
involves integral powers of $\r$ (but see \cite{BS} for
a counterexample). None of the these features is 
essential to the method. The form of the subleading 
terms in the asymptotic expansion is determined by the
bulk field equations.  Notice also that if 
$n$ is not an integer the logarithmic term $\tilde{f}_{(2n)}$ 
in (\ref{ansatz}) would be absent. We assume below that 
$n$ is an integer, since this is the case in all examples
in the literature, but one can easily generalize.

The boundary field $f_{(0)}$ that multiplies the leading behavior, 
$\r^m$, is interpreted as the source for the dual operator.
In the near-boundary analysis one solves the field equations
iteratively by treating the $\r$-variable as a small parameter. 
This yields {\it algebraic} equations for $f_{(2k)}$, $k< n$,
that uniquely determine $f_{(2k)}$ in terms of $f_{(0)}(x)$
and derivatives up to order $2k$. These equations leave 
$f_{(2 n)}(x)$ undetermined\footnote{In the case \label{wif}
$\cf$ is  the metric $G_{\mu \nu}$ or a bulk gauge field $A_\mu$
the bulk field equations partly determine the corresponding 
$f_{(2 n)}(x)$. For instance, as we discussed in the 
previous section, the bulk field equations
determine the divergence and the trace of $g_{(d)ij}(x)$,
but leave undetermined the remaining components.}. 
This was to be expected: the 
coefficient $f_{(
2 n)}(x)$ is the Dirichlet boundary condition
for a solution which is linearly independent from the one 
that starts as $\r^m$. The undetermined function $f_{(2 n)}$
is related to the exact 1-point function of the corresponding 
operator. The logarithmic
term in (\ref{ansatz}) is necessary in order to obtain a
solution. It is related to conformal anomalies of the dual theory, and
it is also fixed in terms of $f_{(0)}(x)$.

To summarize, the asymptotic analysis of the bulk field
equations yields:
\begin{itemize}
\item $f_{(0)}(x)$ is the field theory source,
\item $f_{(2)}(x), ... f_{(2n-2)}$, and $\tilde{f}_{(2n)}$ 
are uniquely determined 
by the bulk field equations and are local functions 
of $f_{(0)}$,
\item $\tilde{f}_{(2n)}$ is related to conformal anomalies,
\item $f_{(2n)}(x)$ is undetermined by the near-boundary analysis.
\end{itemize}

\subsection{Regularization} 
Having obtained the most general asymptotic solution of the 
field equations, we now proceed to compute the on-shell
value of the action. To regularize the on-shell action
we restrict the range of the $\r$ integration, $\r \geq \e$,
and we evaluate the boundary terms at $\r=\e$, where $\e$ is 
a small parameter. A finite number of terms which diverge as 
$\e \rightarrow 0$ can be isolated, so that the on-shell action
takes the form
\be \label{reg}
S_{{\rm reg}}[f_{(0)};\e] =  \int_{\r=\e} \de^{4}x \sqrt{g_{(0)}} 
[\e^{-\nu} a_{(0)}
+\e^{-(\nu+1)} a_{(2)}
+...-\log \e \ a_{(2 \nu)} + \co(\e^0)] \nonumber
\ee
where $\nu$ is a positive number that only depends on the scale
dimension of the dual operator and $a_{(2k)}$ are local functions
of the source(s) $f_{(0)}$. The logarithmic divergence directly
gives the conformal anomaly, as discussed in \cite{HS}. 
The divergences do not depend on $\tilde{f}_{(2n)}$, i.e.
the coefficients that the near-boundary analysis does not 
determine.

\subsection{Counterterms} 
The counterterm action is defined as
\be \label{count}
S_{{\rm ct}}[\cf(x,\e);\e] 
=-{\rm divergent\ terms\ of\ } S_{{\rm reg}}[f_{(0)};\e]
\ee
where divergent terms are expressed in terms of the fields 
$\cf(x,\e)$ `living' at the regulated surface $\r=\e$
and the induced metric there, $\c_{ij} = g_{ij}(x,\e)/\e$. This is 
required
for covariance and entails an ``inversion'' of the expansions 
(\ref{ansatz}) up to the required order. In other words, in order to 
determine $S_{{\rm ct}}$ we first invert the series (\ref{ansatz}) to obtain
$f_{(0)}=f_{(0)}(\cf(x,\e),\e)$, and then substitute in the coefficients
$a_{(2 k)} (f_{(0)}(x)) = a_{(2k)}(\cf(x,\e),\e)$, and finally insert 
those in (\ref{reg}).
 
\subsection{Renormalized on-shell action} 
To obtain the renormalized action we first define a subtracted action 
at the cutoff 
\be \label{renact}
S_{{\rm sub}}[\cf(x,\e);\e] = 
S_{{\rm reg}}[f_{(0)};\e] + S_{{\rm ct}}[\cf(x,\e);\e].
\ee
The subtracted action has
a finite limit as $\e \to 0$, and the renormalized action is a 
functional
of the sources defined by this limit, i.e.
\be \label{sren}
S_{\rm ren}[f_{(0)}] = \lim_{\e \rightarrow 0} S_{{\rm sub}}[\cf;\e]
\ee
The distinction between $S_{\rm sub}$ and $S_{\rm ren}$ is needed 
because
the variations required to obtain correlation functions are performed
before the limit $\e \rightarrow 0$ is taken.

\subsection{Exact 1-point functions} 
The 1-point function of the operator $O_F$ 
in the presence of sources is defined as 
\be \label{1ptdef}
\< O_{F} \>_s = {1 \over \sqrt{g_{(0)}}} 
{\delta S_{\rm{ren}} \over \delta f_{(0)}}
\ee
It can be computed by rewriting it in terms of the fields
living at the regulated boundary,
\be \label{1ptreg}
\< O_{F} \>_s= 
\lim_{\e \to 0} \left( {1 \over \e^{d/2-m}} {1 \over \sqrt{\g}} 
{\delta S_{\rm{sub}} \over \delta F(x,\e)} \right) 
\ee
By construction (\ref{1ptreg})  has a limit
as $\e \to 0$, but it is a good check on all previous
steps to explicitly verify that the divergent terms
indeed cancel. Explicit evaluation of the limit yields  
\be \label{1ptfin}
\< O_{F} \>_s \sim f_{(2n)} + C(f_{(0)}) 
\ee
where $C(f_{(0)})$ is a function that depends
locally on the sources, so it yields contact terms to 
higher point functions. The exact form of $C(f_{(0)})$ depends
on the theory under consideration and in general is scheme
dependent. The coefficient in front of $f_{(2n)}$ 
also depends on the theory under consideration (but it is 
scheme independent). 

\subsection{Ward identities} 
Having obtained explicit formulas for the holographic 1-point functions
it is straightforward to verify whether the expected Ward identities
hold. For instance, the 1-point function of the boundary stress
energy tensor is given by
\be
\< T_{ij} \>_s \sim g_{(d)ij} + C(g_{(0)ij}), 
\ee
see \cite{dHSS, Skenderis:2000in} for the exact formulas.
As discussed in the previous section, the bulk field equations 
only determine the divergence and trace of $g_{(d)ij}$.
This information, however, is enough in order to 
compute the divergence and the trace of $T_{ij}$.
Once Ward identities are established at the level of 
1-point functions (in the presence of sources), they hold in general, since 
$n$-point functions can be obtained by further differentiation
of 1-point functions with respect to the sources.

\subsection{RG transformations} 
The energy scale on the boundary theory is associated with the 
radial coordinate of the bulk spacetime. RG transformations
can be studied by using bulk diffeomorphisms
that induce a Weyl transformation on the boundary metric.
Such transformations have been studied in \cite{ISTY}.
Here we will consider the simplest of such transformations,
\be \label{rresc}
\r = \r' \mu^2, \qquad 
x^i= x^i{}' \mu \ .
\ee
This transformation is an isometry of AdS.
Since we know how bulk fields transform under bulk 
diffeomorphisms, we can readily compute how
the $f_{(2n)}$ transforms under (\ref{rresc}),
and therefore find what is the RG transformation
of $n$-point functions. 

\subsection{$n$-point functions} 
To compute $n$-point functions we need exact 
(as opposed to asymptotic) solutions 
of the bulk field equations with prescribed but 
arbitrary boundary conditions.
Given such an exact solution one can read-off $f_{(2n)}$
as a function of $f_{(0)}$ 
by considering the asymptotics of the solution.
Then $n$-point functions can be computed using 
(\ref{rnpt}).  

Given that the bulk equations are coupled 
non-linear equations, the general Dirichlet 
problem is in general not tractable. 
We proceed by linearizing the bulk field 
equations. Solving the equations for 
linearized fluctuations, i.e. determining
the bulk-to-boundary propagator, allows one
to determine the linear in $f_{(0)}$ term 
of $f_{(2n)}$. This is sufficient in 
order to obtain 2-point functions.
Even in the absence of exact solutions, 
higher point functions can be determined
perturbatively: one solves the bulk field 
equations perturbatively and thus
determines the terms of  $f_{(2n)}$ that are quadratic or
higher orders in $f_{(0)}$.

\section{Example: Massive scalar} \label{msc}

In this section we illustrate the method by 
working through all steps in the simplest possible
example: a free massive scalar field in AdS spacetime.
The action is given by
\be \label{action}
S = \half \int d^{d+1} x \sqrt{G} (G^{\m \n} \pa_\m \F \pa_\n \F 
+ m^2 \F^2)
\ee
The spacetime metric is give by\footnote{
We denote by $\m, \n$, etc. $(d+1)$-dimensional indices and by 
$i,j$, etc. $d$-dimensional indices.}
\be \label{bac}
ds^2 = G_{\m \n} dx^\m d x^\n = 
{d \r^2 \over 4 \r^2} + {1 \over \r}dx^i d x^i.
\ee
The bulk field equation is equal to 
\be \label{feq}
(-\Box_G + m^2) \F =
-{1 \over \sqrt{G}} \pa_\m (\sqrt{G} G^{\m \n} \pa_\n \F) + m^2 \F =0
\ee

\subsection{Asymptotic Solution} 

We want to obtain asymptotic solutions of (\ref{feq}).
The scalar field, however, couples to the Einstein
equation through its stress energy tensor. This 
means that in general we need to solve the coupled
system of gravity-scalar field equations. In 
favorable circumstances the equations decouple 
near the boundary and one can study (\ref{feq}) 
in a fixed gravitational background.
This issue is discussed at length in \cite{holren}
and we refer there for more details. The 
current example is such a favorable case 
(but the example in section \ref{nptsec} is not),
so we proceed by solving (\ref{feq}) in the 
gravitational background given in (\ref{bac}).

We look for a solution of the form
\bea \label{fexp}
\F(x,\r) &=& \r^{(d-\D)/2} \f(x,\r), \nonu 
\f(x,\r) &=& \f_{(0)} + \r \f_{(2)} + \r^2 \f_{(4)} + \cdots
\eea
Inserting this in (\ref{feq}) yields,
\bea \label{feq1}
0=&[&(m^2 - \D (\D -d)) \f(x,\r) \\
&& - \r (\Box_0 \f(x,\r) + 2 (d - 2 \D + 2) \pa_\r \f(x,\r) 
+ 4 \r \pa^2_\r \f(x,\r) )] \nonumber
\eea
where $\Box_0=\d^{ij} \pa_i \pa_j$.
The easiest way to solve (\ref{feq1}) is to successively 
differentiate with respect to $\r$ and then set $\r=0$.
Setting $\r=0$ in (\ref{feq1}) implies,
\be \label{mD}
(m^2 - \D (\D -d))=0
\ee
which is the well-known relation between the mass and the 
conformal weight $\D$ of the dual operator.
With (\ref{mD}) satisfied, (\ref{feq1}) 
reduces to
\be \label{feq2}
\Box_0 \f(x,\r) + 2 (d - 2 \D + 2) \pa_\r \f(x,\r) 
+ 4 \r \pa^2_\r \f(x,\r) =0
\ee
Setting $\r=0$ we get
\be
\f_{(2)}(x) = {1 \over 2 (2 \D - d -2) } \Box_0 \f_{(0)}
\ee
Notice that we solved an {\it algebraic} equation
in order to determine $\f_{(2)}$. In particular, the 
solution remains valid if we change the signature of 
spacetime (the only different in that case is in the 
meaning of $\Box_0$), or we analytically continue the bulk 
metric from AdS to dS. This is a generic feature of the 
asymptotic solutions we discuss. 

Now differentiate (\ref{feq2}) with respect to $\r$ 
and set $\r=0$. The result is
\be
\f_{(4)}(x) = {1 \over 4 (2 \D - d -4) } \Box_0 \f_{(2)}
\ee
Continuing this way one obtains all coefficients 
in the expansion (\ref{fexp}),
\be
\f_{(2 n)} = {1 \over 2n (2 \D - d - 2n)} \Box_0 \f_{(2n-2)}
\ee 

This procedure stops, however, when $2 \D - d - 2n=0$.
In this case we need to introduce a logarithmic term at 
order $\r^{\D/2}$ 
in (\ref{fexp}) to obtain a solution. To be 
concrete consider the case $2 \D -d -2 =0$, i.e.
$\D=d/2 + 1$. The new asymptotic expansion is given by
\be
\f(x,\r) = \f_{(0)} + \r (\f_{(2)} + \log \r \psi_{(2)}) + \cdots
\ee
Inserting this expression in (\ref{feq2}) we now get that 
\be \label{psi2}
\psi_{(2)} = -{1 \over 4} \Box_0 \f_{(0)}
\ee
and we find that $\f_{(2)}$ is {\it not} determined by the field
equations. 

In the general case $\D = d/2 + k$, with $k$ an integer,
a similar computation yields,
\be
\psi_{(2\D-d)} = -{1 \over 2^{2k} \G(k) \G(k+1)} (\Box_0)^k \f_{(0)}
\ee
and again $\f_{(2 \D -d)}$ is {\it not} determined by the 
bulk field equations.

\subsection{Regularization} 
We now evaluate the regularized action on the asymptotic solution
just found,
\bea
S_{{\rm reg}}&=& 
\half \int_{\r \geq \e} d^{d+1} x \sqrt{G} (G^{\m \n} \pa_\m \F \pa_\n \F 
+ m^2 \f^2) \nonu
&=&\half \int_{\r \geq \e} d^{d+1} x \sqrt{G} \F (-\Box_G + m^2) \F 
-\half \int_{\r =\e} d^d x G^{\r \r} \F \pa_\r \F
\eea
where we use the convention to have $\r=\e$ at the lower end of 
radial integration.  Since the bulk field equations are satisfied,
the bulk term vanishes and by inserting the explicit 
asymptotic solution we obtain,
\bea
S_{{\rm reg}}&=& - \int_{\r =\e}  d^d x \e^{-\D +{d \over 2}} \left(
\half (d - \D) \f(x,\e)^2 + \e \f(x,\e) \pa_\e \f(x,\e) \right), \nonu
&=&\int_{\r =\e}  d^d x \left( \e^{-\D +{d \over 2}} a_{(0)} 
+ \e^{-\D +{d \over 2}+1} a_{(2)} + \cdots - \log \e 
a_{(2 \D -d)} \right)
\eea
where 
\bea
&&a_{(0)} = -\half (d-\D) \f_{(0)}^2, \qquad 
a_{(2)} = -(d-\D+1) \f_{(0)} \f_{(2)}=
-{d - \D +1 \over 2(2 \D - d -2)} \f_{(0)} \Box_0 \f_{(0)}, \nonu
&&a_{(2\D -d)}=-{d \over 2^{2k+1} \G(k) \G(k+1)} \f_{(0)} (\Box_0)^k \f_{(0)}
\eea
As promised, the coefficients $a_{(2 \n)}$ of the divergent 
terms are local functions of the source $\f_{(0)}$.
 
\subsection{Counterterms} 
To obtain the counterterms we need to invert the series 
(\ref{fexp}). This is needed because it is $\F(x,\e)$ rather
than $\f_{(0)}$ that transforms as a scalar under
bulk diffeomorphisms at $\r=\e$.
To second order we obtain,
\bea
\f_{(0)} &=& \e^{-(d-\D)/2} \left(\F(x,\e) -
{1 \over 2 (2 \D -d -2)} \Box_\g \F(x,\e) \right) \nonu
\f_{(2)} &=& \e^{-(d-\D)/2-1} {1 \over 2 (2 \D -d -2)} \Box_\g \F(x,\e)
\eea
where $\Box_\g$ is the Laplacian of the induced metric 
$\g_{ij}={1 \over \e} \d_{ij}$ at $\r=\e$. 
These results are sufficient in order to rewrite $a_{(0)}$ 
and $a_{(2)}$ in terms of $\F(x,\e)$. The counterterm action
is then given by (\ref{count})
\be
S_{{\rm ct}} = \int \sqrt{\g} \left(
{d-\D \over 2} \F^2 + 
{1 \over 2 (2 \D -d -2)} \F \Box_\g \F \right) + \cdots
\ee
where the dots indicate higher derivative terms.
Notice that when $\D=d/2+1$ the coefficient of the 
$\F \Box_\g \F$ is replaced by $-{1 \over 4} \log \e$.
Similarly, when $\D=d/2+k$ there is a $k$-derivative
logarithmic counterterm.

\subsection{Renormalized on-shell action} 
The renormalized action in the minimal subtraction scheme
is given by (\ref{renact}). We still have the freedom 
to add finite counterterms. This corresponds to the 
scheme dependence in the field theory. For instance, 
in order to have a manifestly supersymmetric 
scheme where $S_{{\rm ren}}=0$ when evaluated on
the background, it may be necessary to add finite
counterterms. This phenomenon was observed in
\cite{howtogo}.

\subsection{Exact 1-pt function}
Equation (\ref{1ptreg}) adapted to our case gives
\be \label{1ptF}
\< O_{\F} \>_s= 
\lim_{\e \to 0} \left( {1 \over \e^{\D/2}} {1 \over \sqrt{\g}} 
{\delta S_{\rm{sub}} \over \delta \F(x,\e)} \right) 
\ee
For concreteness we will discuss the $\D=d/2+1$ case.
Now,
\bea
\d S_{{\rm sub}} &=& \d S_{{\rm reg}} +  \d S_{{\rm ct}} \nonu
&=& \int_{\r \geq \e} d^{d+1} x \d \F (-\Box_G + m^2) \F \nonu
&&+ \int_{\r=\e} d^d x  \d \F 
\left(-2 \e \pa_\e \F + (d-\D) \F - \half \log \e \Box_\g \F \right)
\eea   
Using the fact that the bulk field equations hold we obtain,
\be
{\d S_{{\rm sub}} \over \delta \F} 
= -2 \e \pa_\e \F + (d-\D) \F - \half \log \e \Box_\g \F
\ee
Inserting this in (\ref{1ptF}) and substituting for 
$\F$ the explicit asymptotic solution we found,
we find that the divergent terms cancel, as they should,
and the finite part is equal to 
\be \label{1ptfi}
\< O_{\F} \>_s= - 2 (\f_{(2)} + \psi_{(2)})
\ee
As promised, the 1-point function depends on the 
part of the asymptotic solution that is not determined
by the near-boundary analysis. 
$\psi_2(x)$ is a local function of the sources, see (\ref{psi2}).
We called such contributions $C(\f_{(0)})$ in (\ref{1ptfin}). 
Actually this term is scheme
dependent. Indeed by adding the finite counterterm
\be \label{anom}
S_{{\rm ct,fin}} = -{1 \over 4} \int d^{d} x \f_{(0)} \Box_0 \f_{(0)}
=- \half \int d^{d} x \sqrt{\g} \ca
\ee
in the action we can remove completely the factor of $\psi_{(2)}$ 
from the the 1-point function. Notice that $\ca$ in (\ref{anom}) 
is the matter conformal anomaly \cite{PS}.

Finally, let us mention that for general $\D$ 
the result is \cite{dHSS}
\be
\< O_{\F} \>_s= - (2 \D -d) \f_{(2\D -d)} + C(\f_{(0)})
\ee

\subsection{RG transformations}
To determine the RG transformations of the correlation
functions we need to determine how the coefficients in the 
asymptotic solution transform under (\ref{rresc}).
Since $\F(x,\r)$ is scalar, we have
\be
\F'(x',\r') = \F(x,\r)
\ee
This equation implies
\bea
\f_{(0)}'(x') &=& \m^{d-\D} \f_{(0)}(x' \m) \label{srg} \\
\f_{(2)}'(x') &=& \m^{d-\D+2} \f_{(2)}(x' \m) \\
\ . \ . &  & \ . \ . \nonu
\psi_{(2 \D-d)}'(x') &=& \m^{\D} \psi_{(2 \D -d)}(x' \m) \\
\f_{(2 \D-d)}'(x') &=& \m^{\D} (\f_{(2 \D -d))}(x' \m) + \log \m^2 
\psi_{(2 \D -d)}(x' \m) )\label{1ptrg}
\eea
Notice that (\ref{srg}) implies 
\be
\m {\pa \over \pa \mu} \f_{(0)}(x' \m)  = -(d - \D) \f_{(0)}(x' \m)
\ee
which is the correct RG transformation rule for a source
of an operator of dimension $\D$.

Now using (\ref{1ptrg}) we can obtain the transformed 1-point
function,
\be \label{ptrg2}
\< O(x') \>_s' = \mu^{\D} \left( \< O(x' \mu ) \>_s 
- (2 \D -d) \log \mu^2 \psi_{(2 \D -d)} (x' \mu) \right) 
\ee
Notice the new term can be obtained by addition
of the following finite counterterm,
\be
S_{{\rm ct,fin}}(\m) = \int d^d x \sqrt{\g} \half \log \mu^2 \ca
\ee
where $\ca$ is the matter conformal anomaly.
This result is as expected: we are computing 
conformal field theory correlation functions.
The correlation function should thus have a 
trivial scale dependence, up to the effects of conformal anomalies.
Indeed, we see from (\ref{ptrg2}) that the transformation of
the 1-point function of the operator of 
dimension $\D$ has the expected scaling term 
and an additional term that is related to the 
conformal anomaly. In other words, all non-trivial 
scale dependence is driven by the conformal anomaly.

\subsection{Correlation functions} 
The considerations so far involved only the near-boundary analysis.
We have derived holographic 1-point functions, but these 
involve coefficients in the asymptotic expansion of the 
bulk fields that the near-boundary analysis does not 
determine. We will now see how these are determined
by obtaining the exact solution of the bulk field equations.
In the case at hand, the field equation is linear in $\F$ 
and can be solved exactly. In more general circumstances
the field equations are non-linear and cannot be solved
in full generality. One may, however, linearize around 
the background and solve the linearized fluctuation 
equations. This is sufficient to obtain 
2-point functions since we only need to know
$\f_{(2 \D -d)}$ to linear order in the source 
in order to obtain them. We will discuss
higher point functions in the next section.

For concreteness we work in $d=4$ and we consider the case 
$\D=d/2+1=3$.
Let $\r=z^2$ and $\F = z^{d/2} \chi$,
and we also Fourier transform in the $x$ coordinates.
The bulk field equation (\ref{feq}) becomes
\be
z^2 \pa_z^2 \chi + z \pa_z \chi - (k^2 z^2 +1) \chi = 0
\ee
This is the modified Bessel differential equation.
The solution that is regular in the interior is
\bea
\chi&=&K_1(k z) \\
&=&{1 \over kz} + \left({1 \over 4} (-1 + 2 \g) -
\half (-\log 2 + \log kz)\right) kz + ... \nonumber
\eea
where in the second line we give the asymptotic 
expansion near $z=0$, and $k=|k|$.

Converting back to the $\r$ coordinate we get
\be
\F(k,\r) = \r^{(d-\D)/2} \f_{(0)}(k)
\left(1 + \r \left(({1 \over 4} (-1 + 2 \g) +
\half \log {k \over 2}) k^2  + 
{1 \over 4} k^2 \log \r  \right) \right) + ...
\ee
where $\f_{(0)}(k)$ represents the overall normalization of $\chi$.
We now read off the various coefficients
\bea
\psi_2(k) &=& {1 \over 4} k^2 \f_{(0)}(k) \to 
\psi_2(x) =- {1 \over 4} \Box \f_{(0)}(x) \label{psie} \\
\phi_{(2)}(k) &=& - 2 \f_{(0)} \left({1 \over 4} (-1 + 2 \g) +
\half \log {k \over 2}\right) k^2 \label{phie}
\eea
Notice that the exact solution correctly reproduces the
value of $\psi_{(2)}$ we determined by the near boundary analysis
(\ref{psi2}). Furthermore, the exact solution determines
$\phi_{(2)}$. Notice that $\phi_{(2)}$ is related 
non-locally to the source $\f_{(0)}$ (their relation involves 
an infinite number of derivatives). 

Inserting in (\ref{1ptfi}) we get
\be
\< O_{\F}(k) \>_s=-2\f_{(0)}(k) \left[
\left({1 \over 4} (-1 + 2 \g) - \half \log 2 + {1 \over 4} k^2\right)
+{1 \over 4} k^2 \log k^2  \right]
\ee
The terms in parenthesis lead to contact terms in the 2-point 
function and can be omitted. We now use (\ref{rnpt})
to obtain the 2-point function,
\bea 
\< O_{\F}(k) O_{\F} (-k) \> 
&=& - {\d \f_{(2)}(k) \over \delta \f_{(0)}(-k)} \nonu
&=&\half k^2 \log {k^2 \over \m^2} \label{2pt}
\eea
where we have also introduced the scale $\m$ (this scale
is introduced by adding a local counterterm proportional 
to the anomaly).

Fourier transforming (\ref{2pt}) we get 
(see appendix A.2 of \cite{difren}).
\be \label{ren2pt1}
\< O_{\F}(x) O_{\F} (0) \> = {4 \over \pi^4} 
\left(-{1 \over 32} \Box \Box {1 \over x^2} \log x^2 M^2 \right)
\ee
where $M=\g \m/2$ and $\g$ is the Euler constant.
The expression in brackets in the right hand side is 
the renormalized version of $1/x^6$ (see (A.1) of \cite{difren}), 
i.e. it is equal to $1/x^6$ for $x \neq 0$, but it is non-singular
(as a distribution) at $x=0$, so 
\be \label{ren2pt}
\< O_{\F}(x) O_{\F} (0) \> = {4 \over \pi^4} \car {1 \over x^6}
\ee
where we used $\car{1 \over x^6} $ to denote  
the renormalized version of $1/x^6$.  This is manifestly the correct 
two-point function for the an operator of dimension 3.
Notice that, as promised, we got the renormalized 2-point 
function. The normalization of (\ref{ren2pt}) also 
agrees with the normalization derived in \cite{FMMR}. 
For $\D=d/2+k$, where $k$ is an integer, the normalization is given by
\be
c_\D = (2 \D - d) {\G(\D) \over \pi^{d/2} \G(\D - {d \over 2})}
\ee
 
\subsection{RG equation} \label{rgeqn}

Let us now consider the RG equation satisfied by 
the two point function we just derived,
\be
M {\pa \over \pa M} \< O_{\F}(x) O_{\F} (0) \>=
{4 \over \pi^4} 
\left(-{1 \over 32} \Box \Box {2 \over x^2} \right) = 
\Box \d^{(4)}(x)
\ee
This is the expected equation. The scale dependence 
of the correlator originates only from the conformal 
anomaly. As derived in \cite{PS} 
(following \cite{osborn}) the trace of the 
stress energy tensor is related to the scale
dependence of the correlator by
\be \label{rg}
\int d^4x \<T^i_i \>  = 
\sum_{k=1}^\infty {1 \over k!} 
\int  d^4 x_1 ... d^4 x_k M {\pa \over \pa M} 
\< O(x_1) ... O(x_k) \>
\ee
Thus local terms in the scale derivative of correlation 
functions lead to conformal anomalies. 
In non-conformal theories,
the $M$ derivative of the correlation function  is non-local 
implying that there is non-trivial $\b$-function. In such 
cases the left hand side of (\ref{rg}) contains a $\b$-function term
as well. We refer to \cite{Erdmenger:2001ja} for related work.

\subsection{$n$-point functions} \label{nptsec}
{}\footnote{The results presented in this section were obtained in collaboration with Dan Freedman and Umut G{\"u}rsoy.}

We discuss in this section the perturbative computation of 
$n$-point functions. We consider 
the case $d=4$ and $\D=d/2+1=3$ and we illustrate the 
method by computing a four-point function.
Our starting point is the following bulk action 
\be \label{agppz}
S = \int d^{5} x \sqrt{G} \left(\half G^{\m \n} \pa_\m \F \pa_\n \F 
+{1 \over 2} m^2 \F^2 - {1 \over 3}  \F^4 \right).
\ee
where $m^2=-3$ in this case.
The reason we choose this specific action is that  
the near-boundary analysis relevant for this action was performed  
in \cite{holren}
in connection with the computation of 2-point functions
in the GPPZ flow \cite{GPPZ}.
We are interested here in computing the 4-point function 
of the operator dual to the scalar in the
AdS vacuum rather than the domain-wall vacuum corresponding to the
GPPZ flow, but as emphasized earlier, the results of the 
near-boundary analysis are valid for {\it any} solution 
of the bulk field equation, so we can freely borrow the 
results of \cite{holren}\footnote{The analysis involves the steps we discussed
in the previous section. In this case, however, one has 
to consider the coupled system of gravity-scalar equations 
rather than study the scalar field equation in a fixed
gravitational background as was done in the previous section.
We refer to section 5.2 of \cite{holren} for details.}. 
The 1-point function is given 
in (5.39) of \cite{holren},
\be \label{1ptint}
\< O_{\F} \>_s= - 2 (\f_{(2)} + \psi_{(2)}) + {2 \over 9} \f_{(0)}^3
\ee 
The last two terms lead to contact terms in correlation functions.
$\psi_{(2)}$ was discussed in the previous section. 
The last term gives an ultra local contribution to the 4-point
function, i.e. it contributes only when all four operators are at the 
same point. In the remainder we discuss the contribution 
at separated points.

The field equation is given by
\be
(-\Box_G + m^2) \F = g \F^3
\ee
where $g=4/3$. 
The idea is now to solve this equation perturbatively in $g$
(one can justify this by 
introducing a coupling constant in the
$\F^4$ term in (\ref{agppz})).
Let 
\be
\F = \F_0 + g \F_1 + \cdots
\ee
then 
\bea
(-\Box_G + m^2) \F_0 &=& 0 \label{0th} \\
(-\Box_G + m^2) \F_1 &=& \F_0^3 \label{1st}
\eea
etc.   
Equation (\ref{0th}) was solved in the previous section. 
It will be convenient to quote the result 
in a way that is valid for other values of $\D$ as well,
\be \label{f0}
\F_0(z,\vec{x}) = \int d^4 y K_\D(z,\vec{x}-\vec{y}) \f_{(0)} (y)
\ee
where the bulk-to-boundary propagator is given by
\be \label{btb}
K_\D(z,\vec{x}-\vec{y}) = C_\D \left({z \over z^2 
+ (\vec{x}-\vec{y})^2}\right)^\D
\ee
and
\be
C_\D = {\G(\D) \over \pi^{d/2} \G(\D-{d \over 2})} 
\ee
Note that $z^2=\r$, as in the previous section. 

Equation (\ref{1st}) is solved by
\be \label{f1}
\F_1(x) = \int d^{d+1} y G_\D (x,y) \F_0(y)^3,
\ee
where $G_\D(x,x')$ is the bulk-to-bulk propagator, 
\be
(-\Box_G + m^2) G_\D(x,x') = \d(x,x'), \qquad 
\d(x,x')={1 \over \sqrt{G}} \d(x-x')
\ee
The explicit solution is given by (see for instance \cite{D'Hoker:2002aw})
\bea \label{prop}
G_\D(x,x')&=& {2^{-\D} C_\D \over 2 \D -d} \xi^\D  
F({\D \over 2},{\D \over 2}+\half;\D-{d \over 2} + 1; \xi^2), \nonu
\xi &=& {2 z z' \over z^2 + z'^2 + (\vec{x} - \vec{x}')^2}
\eea
where  $x=(z,\vec{x})$ and $x=(z',\vec{x}')$, and $F$ is a hypergeometric
function.

Having obtained a solution of the bulk field equation to order $g$,
the next task is to obtain the contribution of $\F_1$ to 
$\f_{(2)}$. From (\ref{f1}) follows that we need the 
near-boundary expansion of the bulk-to-bulk propagator.
The latter is given by
\be \label{nbdry}
G_\D(x,x') = z^\D {1 \over (2 \D -d) } K_\D(z',\vec{x}-\vec{x}') 
+\co(z^{\D+2})
\ee
This follows trivially from (\ref{prop}) upon using 
$F(\D/2,(\D+1)/2;\D-d/2+1; 0)=1$.
Since $\r=z^2$, this exactly has the correct $\r$ dependence
to contribute to 
$\f_{(2)}$ (or $\f_{(2 \D -d)}$ in the general case). 

We are now ready to compute the 4-point function.
By definition,
\be
\<O(x_1) O(x_2) O(x_3) O(x_4) \> = 2 
{\d^3 \f_{(2)} (x_1) \over 
\d \f_{(0)} (x_2) \d \f_{(0)} (x_3) \d \f_{(0)} (x_4)} 
\ee
where we used (\ref{1ptint}). Using (\ref{f1}),  (\ref{nbdry})
and (\ref{f0}) we finally obtain,
\be
\<O(x_1) O(x_2) O(x_3) O(x_4) \> = 3! g 
\int d^{d+1} x \sqrt{G} \prod_{k=1}^4 K_\D (z,(\vec{x}_k-\vec{x})) 
\ee
This is the correctly normalized 4-point function \cite{FMMR}.
The discussion generalizes to different interaction 
terms. The crucial ingredient is the relation (\ref{nbdry}).

\section{RG flows} \label{rgsec}

In the previous section we illustrated in detail how to 
compute holographically renormalized correlation 
functions of conformal field theories. 
The method can be used to obtain correlation functions 
for all quantum field theories that can be obtained
via a deformation or a vev from a CFT that has a holographic dual.

\subsection{The vacuum}

We have seen in section \ref{msc} that the asymptotic expansion
of a scalar field that is dual to a dimension $\D$ operator 
is of the form,
\be
\F=\r^{(d-\D)/2} \f_{(0)} + \cdots + \r^{\D/2} \f_{(2\D-d)} + \cdots
\ee
and that $\f_{(0)}$ has the interpretation of a source
and $\f_{(2\D-d)}$ of a 1-point function. It follows from this 
that if we consider a supergravity solution where the metric 
is asymptotically AdS and there is a non-trivial scalar 
turned on then we either have an operator deformation
of the CFT or the CFT is in a different non-conformal vacuum.

$\bullet$ {\it Operator deformation}. In this case 
the near-boundary expansion of $\F$ is 
$\Phi \sim \r^{(d-\D)/2} \varphi_0$,
and this corresponds to the addition of the term
$\varphi_0 O$ in the Lagrangian of the boundary theory.

$\bullet$ {\it VEV deformation}. In this case 
the near-boundary expansion of $\F$ is 
$\Phi \sim \r^{\D/2} \varphi_0$, and the boundary
Lagrangian is still the same, but the vev of the 
dual operator is non-zero, $\< O \> \sim \varphi_0$,
and the vacuum spontaneously breaks conformal invariance.
 
We now for concreteness restrict ourselves to $d=4$. 
The most general form of a bulk solution that preserves 
Poincar{\'e} invariance in  four dimensions is 
\bea
ds^2 &=& e^{2 A(r)} \d_{ij} dx^i dx^j + dr^2 \nonu
\Phi &=& \Phi(r)
\eea
The action that governs the dynamics of this system is
\be
S=\int d^5 x \sqrt{G} \left( {1 \over 4} R 
+ \half G^{\m \n} \pa_\mu \F \pa_\n \F + V(\F) \right).
\ee

In supergravity theories, the truncation of the theory to 
single scalar usually leads to a potential $V(\F)$ that is related to a 
superpotential $W(\F)$ as
\be \label{potential}
V(\F)={1 \over 2} (\pa_\F W)^2 - {4 \over 3} W^2 .
\ee
This form of the potential, however, also follows from more 
general arguments using gravitational stability. 
A generalized positive energy argument \cite{posthm},
was used in \cite{townsend}, to show that the potential
must have the form (\ref{potential}) when there is a single
scalar field\footnote{For several scalars the obvious generalization
of the form (\ref{potential}) implies stability, but the converse
is not necessarily true.}.  The argument in \cite{townsend}
implies that when the AdS critical point is stable,
there is a ``superpotential'' $W$ such that
the critical point of $V(\F)$ associated with the $AdS$ geometry
is also a critical point of $W$.\footnote{
If one relaxes this  requirement the potential
can always be written in the form (\ref{potential}); one just
views (\ref{potential}) as a differential equation for $W$ \cite{DFGK}.
In this case, however, as the original critical point may not be
a critical point of $W$, the results of \cite{townsend}
about gravitational stability do not necessarily apply.}
In the $AdS/CFT$ correspondence, positivity of energy about a
given $AdS$ critical point is mapped into unitarity of the
corresponding CFT. It follows that in all cases the dual
CFT is unitary the potential can be written as
in (\ref{potential}) \cite{ST}.

When (\ref{potential}) holds a simple $BPS$ analysis \cite{ST,DFGK}
of the domain wall action yields the flow equations 
\be
\label{1storder}
{dA(r)\over dr} =- {2 \over 3} W(\F), \qquad {d \F (r)\over dr} =
\pa_\F W (\F)
\ee
These equations have also been obtained from several other standpoints, 
such as  fermion transformation rules in the (truncated) 
supergravity theory \cite{FGPW},
and the Hamilton-Jacobi framework \cite{dBVV,deBoer:2000cz}. 

\subsection{Correlation functions}

Solutions to the first order equations (\ref{1storder}) 
provide the vacuum of the dual quantum field theory. 
Many such classical solutions are known (see, for instance, 
\cite{FGPW,FGPW2,bs,GPPZ,pstarinets,pz0002172,warner,BS}).
The computation of correlation function along RG flows
is analogous to the case of correlators of a CFT 
we just discussed. The details have been spelled out in 
\cite{howtogo} and \cite{holren} to which we refer for a more 
complete discussion. Here we will only discuss a few 
selected topics.

The first step in computing correlation functions
is to perform the near-boundary analysis. At the end 
of this analysis one ends up with a number of 
exact 1-point functions and a number of Ward identities 
that relates them. As emphasized on many occasions, the 
results of this step only depend on the action 
one starts from and can be applied to any particular 
solution.

Let us consider the case of a scalar coupled to gravity.
This system is relevant for the study of correlation 
functions of the stress energy tensor and a scalar operator.
The near-boundary analysis in this case leads to the following 
diffeomorphism and Weyl Ward identities,
\bea
\nabla^i \<T_{ij} \>_s &=& - \<O\>_s \nabla_j \phi_{(0)} \label{diffeo} \\
\<T_{\ i}^i \>_s &=& (\D-4) \phi_{(0)} \<O\>_s + \ca   \label{trace}
\eea
where $\ca$ is the holographic conformal anomaly.
These results hold provided the bulk action is covariant and admits
an AdS critical point. It is a good check on computations
of the near-boundary analysis to verify that 
the actual 1-point functions do satisfy these Ward identities.
It is also easy to establish that these are the expected field theory  
Ward identities. Indeed by definition,
\be \label{variation}
\d S_{{\rm ren}} = \int d^4 x \sqrt{g_{(0)}} 
[\half \<T_{ij}\> \delta g_{(0)}^{ij}
+ \<O \> \delta \f_{(0)}]
\ee
Invariance of (\ref{variation}) under diffeomorphisms
\be \label{diff4}
\delta g_{(0)}^{ij} = - (\nabla^i \xi^j + \nabla^j \xi^i), \qquad
\delta \f_{(0)} = \x^i \nabla_i \f_{(0)}
\ee
yields (\ref{diffeo}) and invariance, up to an anomaly, under
Weyl transformations 
\be \label{weyl1}
\delta g_{(0)}^{ij} = - 2 \s g_{(0)}^{ij}, \qquad
\delta \f_{(0)}=(\D - 4) \s \f_{(0)}
\ee
yields (\ref{trace}).

Let us now use these results in the context of RG flows.

$\bullet$ {\it Operator deformation} \newline
In this case there is a non-zero background value of $\phi_0=\varphi_0$, so 
we obtain from (\ref{trace}),
\be
\<T_{\ i}^i \> = (\D-4) \varphi_0 \<O\> + \ca 
\ee
which leads to the identification of the $\b$ function 
\be
\b=(\D-4) \varphi_0
\ee
Notice that this is the $\b$-function for the coupling constant 
associated with the operator we added in the Lagrangian, not
the $\b$-function for the gauge coupling constant.
Thus we get the correct RG equation,
\be
T_{\ i}^i = \b O + \ca,
\ee
see the discussion in section \ref{rgeqn}.

$\bullet$ {\it VEV deformation} \newline
In this case it is $\< O \>$ which has a background value,
$\< O \>_B$.
In this case combining (\ref{diffeo}) and (\ref{trace})
and going to momentum space we derive for the connected 
correlator\footnote{Notice that the stress energy tensor
as defined by (\ref{variation}) includes the term
$g_{(0) ij} \phi_{(0)} O$. This originate from the term
$\int \sqrt{g_{(0)}} \phi_{(0)} O$. 
The standard field theory stress energy 
tensor does not include this term and it
is related to $T_{ij}$ by
$\<T_{ij}\>_{QFT} = \<T_{ij}\> + \varphi_{(0)} \<O\> g_{(0)ij}$,
where $\varphi_{(0)}$ is the fluctuation part of the source $\phi_{(0)}$. 
When the background vev $\<O\>$ does not vanish 
one must correct for this effect. This has been taken into 
account in (\ref{2ptsp}).}
\be \label{2ptsp}
\<T_{ij}(p) O (-p)\> = -{{\D-4} \over 3} \< O \>_B \p_{ij}
\ee
where $\pi_{ij} =\delta_{ij} -p_ip_j/p^2$. 
We thus find that the two-point function exhibits a 
massless pole. This is the expected dilaton pole due to 
the Goldstone boson of spontaneously broken 
conformal symmetry.

It remains to actually compute the correlators. As long as 
the derivation is consistent with the near-boundary analysis, 
the correlation functions are guaranteed to exhibit the 
physics of operator deformation or spontaneous
symmetry breaking we just discussed.

To obtain 2-point function we need to linearize around the 
domain-wall background and then solve the resulting fluctuation 
equations. The treatment of fluctuations is universal for
the gravity-scalar-vector sector \cite{Notes,Anatomy,bs2}. 
What is important for the computation of the correlation
function is that one should consider linear combinations
of fluctuations that only depend
on the conformal structure at infinity, i.e.
the fluctuations should not change if we change the representative
of the conformal structure \cite{howtogo}.
To illustrate this point we discuss the fluctuation
equations of the gravity-scalar sectors.

We look for fluctuations around the solution $(A(r), \vf_B(r))$
of (\ref{1storder}). The background plus linear fluctuations is
described by
\bea
ds^2&=& e^{2A(r)}[\d_{ij} + h_{ij}(x,r)] dx^i dx^j + (1+ h_{rr}) dr^2
\nonu
\Phi&=& \vf_B(r) + \tvf(x,r)
\eea
where $h_{ij}$, $h_{rr}$ and $\tvf$ are considered infinitesimal.
This choice does not completely fix the bulk diffeomorphisms.
One can perform the one-parameter family of `gauge transformations'
\be \label{diffeomo}
r= r' + \e^r(r',x'), \qquad
x^i = x'^i + \e^i(r',x')
\ee
with
\be \label{edif}
\e^i = \d^{ij} \int_r^\infty dr' e^{-2 A(r)}\pa_j \e^r
\ee
where we only display the fluctuation-independent part of $\e^i$.
These diffeos are related to those which induce the Weyl transformation
(\ref{weyl1}) of the sources. The gauge choice is also 
left invariant by the linearization of the $4d$ diffeomorphisms in 
(\ref{diff4}).

We decompose the metric fluctuation as
\be
h_{ij}(x,r) = h^{T}_{ij}(x,r) + \d_{ij} {1 \over 4} h(x,r)
- \pa_i \pa_j H(x,r) + \nabla_{(i} h_{i)}^L
\ee
$4d$ diffeos can be used to set $h_{i}^L=0$ and we choose to do so.
The equation for the transverse traceless modes decouples from the
equations for $(\tvf, h, H, h_{rr})$.
The coupled graviton-scalar field equations in the axial gauge 
where $h_{rr}=0$
were derived in \cite{Notes}, and we now include $h_{rr}$. The fluctuation
equations are \cite{howtogo}
\bea
&&[\pa_r^2 + 4 A' \pa_r + e^{-2A} \Box] f(x,r)=0, \ \
h^T_{ij} = h^T_{(0)ij} f(x,r) \label{tteqn} \\
&&h' = - {16 \over 3} \vf' \tvf + 4 A' h_{rr} \label{heqn1} \\
&&H'' + 4 A' H' - \half e^{-2A} h - h_{rr} e^{-2 A} =0 \\
&& 2 A' H' = \half e^{-2 A} h + {8 \over 3}{1 \over p^2} W_{\varphi}
(\tvf' - W_{\varphi\varphi} \tvf - \half W_{\varphi} h_{rr})
\eea
where $h^T_{(0)ij}$ is transverse, traceless, and independent of $r$.

The fluctuations $(\tvf, h, H, h_{rr})$ transform under the
`gauge transformations' in (\ref{diffeo}). This implies that they 
depend not only on the conformal structure at infinity,
but also on the specific representative chosen. To remedy for this we 
look for gauge invariant combinations.
Such combinations are the following \cite{howtogo},
\be \label{ginv}
R\equiv h_{rr} - 2 \pa_r \left({\tvf \over W_{\varphi}}\right), \qquad
h+{16 \over 3} {W \over W_{\varphi}} \tvf,
\qquad H'- {2 \over W_{\varphi}} e^{-2A} \tvf
\ee
In terms of these variables the equations simplify, 
\bea \label{eqnsim}
&&h+{16 \over 3} {W \over W_{\varphi}} \tvf =
- {16 \over 3} {e^{2A} \over p^2}\left(
R (W W_{\varphi\varphi} - {4 \over 3} W^2 - \half W_{\varphi}^2) +\half R' W \right) \label{hfeqn}\\
&&H'- {2 \over W_{\varphi}} e^{-2A} \tvf = {1 \over p^2}
\left(2 R (W_{\varphi\varphi} - {4 \over 3} W) + R' \right)
\eea

Equation (\ref{heqn1}) takes the form
\be \label{eqh2}
\left(h+{16 \over 3} {W \over W_{\varphi}} \tvf \right)' = -{8 \over 3} W R
\ee
Differentiating (\ref{hfeqn}) leads to the
second order differential equation
\be \label{ODE}
R''+(2W_{\varphi\varphi}-4W) R' -(4 W_{\varphi}^2 - 2 W_{\varphi}
W_{\varphi\varphi\varphi} - {32\over 9} W^2
+{8\over 3} W W_{\varphi\varphi}
+ p^2 e^{-2A}) R = 0
\ee

We thus find that in order to obtain 2-point functions we need to solve
the second order ODE (\ref{ODE}). In favorable circumstances
this ODE reduces to a hypergeometric equation whose solutions 
and their asymptotics are known. In such cases one can explicitly 
work out the 2-point functions. This was explicitly done for 
two RG flows, one involving an operator deformation and the other 
a vev in \cite{howtogo}. Equivalent results were obtained simultaneously in 
\cite{Muck:2001cy} and earlier progress was reported in 
\cite{Arutyunov:2000rq}. We refer to these works for the details.

\section{Conclusions}

In these lectures notes we presented a systematic method for 
computing renormalized correlation functions in the 
gravity/gauge theory correspondence. The method is complete 
and can be applied in all cases where the spacetime is asymptotically 
AdS.\footnote{The method has been developed in the Lagrangian formalism.
One can also recast holographic RG flows in the Hamilton-Jacobi
formalism \cite{dBVV,deBoer:2000cz}. By exploring the 
relation between the Lagrangian and Hamilton-Jacobi formalisms,
one can transcribe all steps of the 
holographic renormalization method 
to the Hamilton-Jacobi (HJ) method. The issue
of renormalization in this  setting 
has been addressed in \cite{Martelli:2002sp}.}
This means that the dual theory should 
flow in the UV to a fixed point. Even though this 
appears to be very general there is a class of gravity/gauge theory
dualities where the spacetime is not asymptotically AdS.

The first such example is the case of the gravity/gauge theory 
duality involving non-conformal branes 
\cite{Itzhaki:1998dd,Boonstra:1998mp}. In this case the 
duality involves spacetimes that are conformal 
to AdS. It should be possible to 
generalize the method by working in the dual frame, 
i.e. the frame where the spacetime is exactly AdS.
It would be interesting to work out the details.

Another case of interest is the dualities of the 
Klebanov-Strassler type \cite{Klebanov:2000hb}.
A computation of a two-point
function for this geometry has been presented in
\cite{Krasnitz:2000ir}. The results 
of that paper indicate that a generalization 
of the method to these geometries should be 
possible. The main difference of this case with the 
cases discussed here is that the
spacetime geometry contains logarithms at 
leading order. This means that the form of the 
asymptotics in the near-boundary analysis 
should be modified accordingly. 

It would also be interesting to develop the 
method from the ten dimensional point of view.
Our discussion was entirely in terms 
of $d+1$-dimensional fields. There is no problem 
of principle with this since we can always KK
reduce the $10d$ theory to $(d+1)$-dimensions.
The $10d$ perspective, however, can have 
many advantages. To name a few: geometries that are singular
from the $5d$ point of view may be
non-singular from the $10d$ point of view;
each of the KK towers is represented by 
a single field, and more importantly 
some of the gravity/gauge theory dualities,
such as the Polchinski-Strassler duality \cite{Polchinski:2000uf}, 
are formulated more naturally from the $10d$ point of view. On the other 
hand, the $10d$ geometries of the form $AdS \times  M$ have 
degenerate boundaries, the asymptotic expansions are not
universal, and the subtractions are made by counterterms that 
look non-covariant from the $10d$ point of view \cite{m1,m2,m3}.
A higher dimensional point of view is also relevant for 
extending the program to cover dualities such as \cite{Maldacena:2000yy}.

In general, fully understanding how to deal with degenerate boundaries
would most likely lead to developing tools that are 
useful in extending the program of holographic renormalization 
to more general geometries. 

Another future direction is to apply the holographic renormalization
technique to the DBI action. This would be relevant for computing 
correlation functions in the defect RG 
flows \cite{Skenderis:2002vf,Karch:2002sh,Yamaguchi:2002pa} 
that arise in the context of the 
AdS/dCFT duality \cite{Karch:2000gx}.
Relevant work can be found in \cite{Graham:1999pm}. 
Finally, it would also be interesting to 
investigate the generalization of the method to the plane wave 
geometry.

\section*{Acknowledgments} I would like to thank
Dan Freedman and Umut G{\"u}rsoy for collaboration
on unpublished work presented in section \ref{nptsec},
and Marika Taylor for her comments on the manuscript.
This research is supported in part by the
National Science Foundation grant PHY-9802484.

\appendix

\section{Analytic continuation to De Sitter}

Recently there has been interest in a possible holographic
interpretation of dS gravity \cite{Witten:2001kn,Strominger:2001pn}.
Looking for supporting evidence to a possible dS/CFT correspondence,
many authors investigated the asymptotic symmetries of 
asymptotically de Sitter spacetime and the associated conserved charges 
\cite{Strominger:2001pn,Mazur:2001aa,Nojiri:2001mf,Klemm:2001ea, Spradlin:2001pw,BBM}.
We show in this appendix that
the near-boundary analysis of asymptotically AdS spacetimes
can be analytically continued 
to asymptotically dS spacetimes. It follows that counterterms and exact 
1-point functions in the presence of sources have 
a straightforward continuation. A particular case 
is the holographic stress energy tensor.
Thus, although the results on the asymptotic symmetries
are consistent with a possible dS/CFT correspondence,
they do not neccessarily constitute a
piece of evidence for such a correspondence
since they follow from corresponding 
AdS results and their holographic interpretation
may be due to the AdS/CFT duality.
We emphasize that this discussion is valid 
for local properties only. The global properties
of de Sitter spacetime differ significantly from 
the global properties of AdS spacetimes. To compute 
correlation functions using the methods described
in this review, one has to address global issues 
as well. Exact solutions for massive scalar fields
on de Sitter spacetimes have been recently discussed
in this context in \cite{Spradlin:2001nb,Bousso:2001mw}. 

Recall that Einstein's equations in the two cases read
(our curvature conventions are given in footnote \ref{conventions}),
\bea
R_{\mu \nu} = {d \over l_{AdS}^2} G_{\mu \nu} &&
\qquad \qquad {\rm AdS} \\
R_{\mu \nu} = - {d \over l_{dS}^2} G_{\mu \nu} && \qquad \qquad {\rm dS} 
\eea
So the one equation is mapped to the other by 
\be \label{l}
l_{AdS}^2 \leftrightarrow - l_{dS}^2 
\ee

Let us now consider the metrics. The Euclidean AdS metric in Poincar{\'e}
coordinates,
\be
ds_{AdS}^2 = {l_{AdS}^2 \over r^2} (dr^2 + dx^i d x^i),
\ee
is mapped to the ``big bang'' metric,
\be \label{dsBB}
ds_{dS}^2 = {l_{dS}^2 \over t^2} (-dt^2 + dx^i d x^i),
\ee
by using (\ref{l}) and in addition take
\be \label{r}
r^2 \to -t^2.
\ee
Notice that the de Sitter spacetime has two boundaries,
one at past infinity, and another at future infinity.
The metric (\ref{dsBB}) covers only half of de Sitter 
spacetime, and only one of the two boundaries is covered by this 
coordinate patch.

Inspection of the arguments presented in section \ref{aads} shows that 
near each of its boundaries, an asymptotically dS metric 
can be brought to the form,
\bea \label{dsm}
ds_{dS}^2 &=& l_{ds}^2 \left(-{d \tilde{\r}^2 \over 4 \tilde{\r}^2}
+ {1 \over \tilde{\r}} \tilde{g}_{ij}(x,\tilde{\r})  dx^i d x^j \right) \nonu
\tilde{g}_{ij}(x,\tilde{\r})&=&\tilde{g}_{(0) ij} + \tilde{\r} 
\tilde{g}_{(2) ij} + \cdots
\eea
where $\tilde{\r}$ is related to $t$ by $\tilde{\r}=t^2$.
As we discussed in section \ref{aads}, the near-boundary 
equations for the coefficients $g_{(k) ij}$ are algebraic.
It follows that one can immediately write the 
solutions for $\tilde{g}_{(k) ij}$ starting from the solutions
$g_{(k) ij}$ relevant for the asymptotically AdS space,
\bea \label{tg}
\tilde{g}_{(4 k) ij}  &=&  g_{(4 k) ij} \\
\tilde{g}_{(4k+2) ij}  &=& - g_{(4k+2) ij}, \nonumber
\eea
where $k=0,1,...$
The minus sign originates from (\ref{r}), which implies 
$\r \to - \tilde{\r}$. 

Notice that there are two asymptotic expansions, one for each boundary. 
These expansions should be matched in the region of overlap
of the two coordinate patches they cover.
Given that the field equations are second order in $\tilde{\r}$,
if we fix $\tilde{g}_{(0)}$ at both boundaries, we effectively
uniquely fix the solution throughout spacetime. The coefficients
$\tilde{g}_{(d)ij}(x)$ that the near-boundary analysis does not 
determine are now determined by matching the two asymptotic 
expansions. Recall that correlation functions in the AdS/CFT 
correspondence are determined by obtaining global solutions,
and reading off the coefficient $\tilde{g}_{(d)ij}$. 
The above argument suggests that the $\tilde{g}_{(d)ij}(x)$
is a local function of the sources, and thus the corresponding 
correlation functions consist of only contact terms.
It would be interesting to make this argument more precise
and to also investigate different boundary conditions.
For a closely related discussion we refer to \cite{Balasubramanian:2002zh}.
For now we follow the practice in most of current literature and 
focus on one of the two boundaries of dS.

Given the asymptotic solution one can proceed to renormalize the 
on-shell action. We will present in parallel both the 
AdS and dS case. The action is given by
\be \label{actionEH}
S={1 \over 16 \p G}[\int_{M} d^{d+1}x\, 
\sqrt{|G|}\, (R + 2 \L) 
- \int_{\pa M} d^d x\, \sqrt{|\c|}\, 2 K],
\ee
where $K$ is the trace of the second fundamental form and
$\c$ is the induced metric on the boundary. 

The (regularized) on-shell value of the bulk term in (\ref{actionEH}) 
can be computed using the following results.
From Einstein equations we get,
\bea
{\rm dS:} && R + 2 \L = -{2 d \over l_{dS}^2} \nonu
{\rm AdS:} &&  R + 2 \L =  {2 d \over l_{AdS}^2} 
\eea
Furthermore we need the trace of the second fundamental form
\bea
{\rm dS:} && K = -{1 \over l_{dS}^2} (d - \r \Tr g^{-1} \partial_\r g) \nonu
{\rm AdS:} &&  K = {1 \over l_{AdS}^2} 
(d - \tilde{\r} \Tr \tilde{g}^{-1} \partial_{\tilde{\r}} \tilde{g}) 
\eea
where the relative minus sign in the dS case compared to AdS 
case originates from the fact that the normal to the 
boundary vector is spacelike in the one case and 
timelike in the other. 

Inserting these values in (\ref{actionEH}) we get
\bea
S_{AdS} &=& {l_{AdS}^{d-1} \over 16 \pi G}
\left(\int d^d x \left[\int_\e \left(
d \r {d \sqrt{\det g(x,\r)} \over \r^{d/2+1}}\right)
- 2 {\sqrt{\det g(x,\e)} \over \e^{d/2}}
(d- \e \Tr g^{-1} \pa_\e g)  \right] \right) \nonu 
S_{dS} &=& - {l_{dS}^{d-1} \over 16 \pi G}
\left(\int d^d x \left[\int_{\tilde{\e}} \left(d \tilde{\r} 
{d \sqrt{\det \tilde{g}(x,\tilde{\r})} 
\over \tilde{\r}^{d/2+1}}\right)
- 2 {\sqrt{\det \tilde{g}(x,\e)} \over \tilde{\e}^{d/2}}
(d-\tilde{\e} \Tr \tilde{g}^{-1} \pa_{\tilde{\e}} \tilde{g})
\right] \right) \nonumber
\eea
Thus the regularized dS on-shell action can be obtained from
the AdS one by simply multiplying by minus ones
and taking $g \to \tilde{g}$.
It follows that the corresponding counterterms required to render the 
action finite can also be obtained from the AdS ones
by $g \to \tilde{g}$ and by multiplying by minus one.

We now want to compare the Brown-York stress energy tensor \cite{BrownYork}
in the AdS and dS cases. By definition,
\be \label{tij}
T_{ij} = {2 \over \sqrt{g_{(0)}}} 
{\delta S_{{\rm ren}} \over \d g_{(0)}^{ij}}
\ee
where $S_{{\rm ren}}$ is defined as usual (\ref{renact}).
The computation of (\ref{tij}) was carried out in full generality 
in \cite{dHSS}. Since the renormalized actions of dS and 
AdS differ only by sign, the results carry over to the dS case 
as well. One simply has to multiply by an overall minus
and change  $g_{(k)}$ to $\tilde{g}_{(k)}$. Let us discuss 
in some detail the three and five dimensional cases.
All other cases are similar.

$\bullet$ {\it d+1=3 case} \newline
The relevant equation is (3.10) of \cite{dHSS}.
We get 
\be
T^{dS}_{ij} = - {l_{dS} \over 8 \pi G} 
(\tilde{g}_{(2)ij} - \tilde{g}_{(0)ij} \Tr \tilde{g}_{(2)} )
={l_{dS} \over 8 \pi G}
(g_{(2)ij} - g_{(0)ij} \Tr g_{(2)} )
\ee
where we used (\ref{tg}). Thus,
\be
T_{ij}^{dS}=T_{ij}^{AdS}
\ee
and 
\be
T^i_i = -{1 \over 24} \left({3 l_{dS} \over 2 G} \right) R.
\ee
Thus the central charge $c=3  l_{dS}/2 G$ comes out positive.

$\bullet$ {\it d+1=5 case} \newline
In this case the relevant equation is (3.15) of \cite{dHSS},
and we get
\be
T_{ij}=-{4 l_{dS} ^3 \over 16 \p \GN} [\tilde{g}_{(4)ij}
-{1 \over 8} \tilde{g}_{(0)ij} 
[(\Tr\, \tilde{g}_{(2)})^2-\Tr\, \tilde{g}_{(2)}^2] -
\half (\tilde{g}_{(2)}^2)_{ij} + {1 \over 4} \tilde{g}_{(2)ij} 
\Tr\, \tilde{g}_{(2)}]
\ee
This expression involves $\tilde{g}_{(4)}$ or $\tilde{g}_{(2)}^2$.
It follows from (\ref{tg})
that the dS stress energy tensor is minus the 
AdS one,
\be \label{4dtij}
T_{ij}^{dS} = - T_{ij}^{AdS}
\ee
In asymptotically AdS spacetimes positive energy theorems \cite{Gibbons:jg} 
guarantee that that the mass of global AdS spacetime is the lowest 
possible mass. In the AdS/CFT correspondence, this translates
to the statement the ground state energy is the lowest energy of the system.
Provided there are no subtleties in converting the local statement 
(\ref{4dtij}) to the statement about the corresponding 
charges, the relation (\ref{4dtij}) implies that the dS ``mass'' 
as defined in \cite{BBM} is the 
maximum possible mass among all asymptotically de Sitter 
spacetimes\footnote{One also needs to establish 
that the analytic continuation maps the ``allowed'' 
asymptotically dS solutions (i.e. without cosmological singularities) 
to well-behaved asymptotically AdS solutions (i.e. no naked singularities).}.
This was conjectured to be true in \cite{BBM}.

\end{document}